\input{psfig}
\input{epsf}
\documentstyle[12pt]{article}

\def\Journal#1#2#3#4{{#1} {\bf #2}, #3 (#4)}

\def\NPB{{\em Nucl.~Phys.}~B}
\def\PLB{{\em Phys.~Lett.}~B}
\def\PRL{\em Phys.~Rev.~Lett.}
\def\PRD{{\em Phys.~Rev.}~D}

\def\be{\begin{equation}}
\def\ee{\end{equation}}
\def\bea{\begin{eqnarray}}
\def\eea{\end{eqnarray}}
\def\ba{\begin{array}}
\def\ea{\end{array}}
\def\nn{\nonumber}
\def\simge{\mathrel{%
   \rlap{\raise 0.511ex \hbox{$>$}}{\lower 0.511ex \hbox{$\sim$}}}}
\def\simle{\mathrel{
   \rlap{\raise 0.511ex \hbox{$<$}}{\lower 0.511ex \hbox{$\sim$}}}}
 
\def\slashchar#1{\setbox0=\hbox{$#1$}           
   \dimen0=\wd0                                 
   \setbox1=\hbox{/} \dimen1=\wd1               
   \ifdim\dimen0>\dimen1                        
      \rlap{\hbox to \dimen0{\hfil/\hfil}}      
      #1                                        
   \else                                        
      \rlap{\hbox to \dimen1{\hfil$#1$\hfil}}   
      /                                         
   \fi}                                         %
\def\ts{\thinspace}

\def\ra{\rightarrow}

\def\ol{\bar}

\def\CA{{\cal A}}
\def\CB{{\cal B}}
\def\CC{{\cal C}}
\def\CD{{\cal D}}

\def\CL{{\cal L}}
\def\CM{{\cal M}}

\def\ecm{\sqrt{s}}

\def\atro{\alpha_{\rho_T}}
\def\Few{F_\pi}
\def\Ntc{N_{TC}}

\def\sutc{SU(\Ntc)}

\def\thw{\theta_W}
\def\kslash{\raise.15ex\hbox{/}\kern-.57em k}
\def\tro{\rho_{T}}

\def\troz{\rho_{T}^0}
\def\tropm{\rho_{T}^\pm}

\def\tom{\omega_T}
\def\tpi{\pi_T}
\def\tpipm{\pi_T^\pm}
\def\tpimp{\pi_T^\mp}

\def\tpiz{\pi_T^0}
\def\tpipr{\pi_T^{0 \ts\prime}}

\def\mm{\mu^+\mu^-}

\def\mev{{\rm MeV}}
\def\gev{{\rm GeV}}
\def\tev{{\rm TeV}}

\def\nb{{\rm nb}}

\def\pb{{\rm pb}}
\def\ipb{{\rm pb}^{-1}}

\def\ifb{{\rm fb}^{-1}}

\def\third{{\textstyle{ { 1\over { 3 } }}}}

\begin{document}
\title{
\vskip -15mm
\begin{flushright}
\vskip -15mm
{\small FERMILAB-Pub-98/065-T\\
BUHEP-98-2\\
hep-ph/9802368\\}
\vskip 5mm
\end{flushright}
{\Large{\bf \hskip 0.38truein
Narrow Technihadron Production \hfil\break at the First Muon Collider}}\\
}
\author{
\centerline{{\small Estia Eichten$^{1}$\thanks{eichten@fnal.gov},}
{\small Kenneth Lane$^{2}$\thanks{lane@buphyc.bu.edu}, and}
{\small John Womersley$^{1}$\thanks{womersley@fnal.gov}}}\\
\centerline{{\small {$^{1}$}Fermilab, P.O.~Box 500, Batavia, IL 60510}}\\
\centerline{{\small {$^{2}$}Department of Physics, Boston University, 
590 Commonwealth Avenue, Boston, MA 02215}}\\
}
\maketitle
\begin{abstract}
In modern technicolor models, there exist very narrow spin-zero and
spin-one neutral technihadrons---$\tpiz$, $\troz$ and $\tom$---with
masses of a few $100\,\gev$. The large coupling of $\tpiz$ to
$\mu^+\mu^-$, the direct coupling of $\troz$ and $\tom$ to the photon
and $Z^0$, and the superb energy resolution of the First Muon Collider
may make it possible to resolve these technihadrons and produce them at
extraordinarily large rates.
\end{abstract}


\newpage

The next big step in collider physics after the Large Hadron Collider is a
matter of great importance and considerable debate. Electron-positron
linear colliders with center-of-mass energy $\ecm = 500$--$1000\,\gev$ are
touted for the clean environment of their interaction region and high
signal-to-background rates. Hadron colliders, with $pp$ or $\ol p p$ beams,
can make a substantial leap beyond the LHC with $\ecm \simge 100\,\tev$ and
integrated luminosities exceeding $100\,\ifb$ per year (hence subprocess
energies exceeding $10\,\tev$). The proponents of $\mm$ colliders claim
they can deliver the the best aspects of both: relatively clean and
background-free collisions (at least for $\vert \cos\theta\vert \simle
0.95$) and very high collision energies, in the range~2--$4\,\tev$.
However, the potential difficulties of a muon collider are so great that a
successful low-energy prototype, the First Muon Collider (FMC) with $\ecm =
100$--$500\,\gev$, certainly must be demonstrated.

So far, the primary justification for a low-energy muon collider has been
copious resonant production of neutral Higgs bosons, $H^0$, such as
expected in minimal or multi-Higgs doublet standard models or their
supersymmetric variants. Because the $H^0$ coupling to $\mm$ is of order
$m_\mu/v$, where $v = 246\,\gev$, the Higgs cross section is $(m_{\mu}/m_e)^2
= 10^4$ times greater in the FMC than it is in an $e^+e^-$ collider.
Furthermore, the beam momentum resolution claimed for the FMC, $\delta p/p
= 10^{-5}$--$10^{-3}$~\cite{deltap}, is much better than can be achieved in
linear $e^+e^-$ colliders, making $\mm$ production rates even larger.
Although neutral Higgs bosons will be discovered at the Tevatron or LHC,
the advantages that a muon collider has over a hadron collider for studying
the details of $H^0$ production and decay are obvious.

In this letter we point out another strong motivation for the First Muon
Collider: Modern technicolor models, particularly topcolor-assisted
technicolor (TC2)~\cite{tctwo} with a walking gauge coupling~\cite{wtc},
are expected to contain many technihadron states, some lying at the low
energies the FMC will probe. These states, specifically, neutral
technipions and technivectors, are very narrow and can be produced as
$s$-channel resonances in $\mm$ annihilation. The cross sections on
resonance are enormous---from 1/10 to 10 nanobarns. The energy resolution
of the FMC permits a substantial part of these peak production rates to be
realized. In no other machine can such precise and spectactular studies of
low-mass technihadrons be executed.
~\footnote{The lightest
technihadrons should be accessible at the Tevatron collider in Run~II
or~III~\cite{tpitev}. They are easily produced and detected at the LHC at
moderate luminosities.}

We assume that the technicolor gauge group is $\sutc$ and take $\Ntc =4$ in
calculations. Its walking gauge coupling is achieved by a large number of
isodoublets of technifermions transforming according to the fundamental
representation of $\sutc$. We consider the phenomenology of only the
lightest color-singlet, spin-zero and one technihadrons and assume that
they may be considered in isolation for a {\it limited} range of the $\mm$
energy $\ecm$ about their masses.~\footnote{Technicolor models with
QCD-like dynamics are incompatible with precision electroweak
measurements~\cite{pettests}, but these proofs are inapplicable to walking
technicolor, principally because the electroweak spectral functions cannot
be saturated by a single vector and axial vector resonance~\cite{glasgow}.
Also see Ref.~\cite{edr}.}
These technihadrons consist of a single isotriplet and
isosinglet of vectors, $\troz$, $\tropm$ and $\tom$, and pseudoscalars
$\tpiz$, $\tpipm$, and $\tpipr$. The latter are in addition to the
longitudinal weak bosons, $W^\pm_L$ and $Z^0_L$, which are technipion bound
states of all the technifermions. In TC2 there is no need for large
technifermion isospin splitting associated with the top-bottom mass
difference. Thus, the lightest $\tro$ and $\tom$ are approximately
degenerate. The lightest charged and neutral technipions also should have
roughly the same mass, but there may be appreciable $\tpiz$--$\tpipr$
mixing. If that happens, the lightest neutral technipions are really
techni-$\ol U U$ and $\ol D D$ bound states. Finally, for purposes of
discussing signals at the FMC, we take the lightest technihadron masses to
be $M_{\tro} \cong M_{\tom} \sim 200\,\gev$; $M_{\tpi} \sim 100\,\gev$.

Technipion decays are induced mainly by extended technicolor (ETC)
interactions which couple them to quarks and leptons~\cite{etceekl}. These
couplings are Higgs-like, and so technipions are expected to decay into the
heaviest fermion pairs allowed. In TC2, only a few GeV of the top-quark's
mass is generated by ETC, so there is no great preference for $\tpi$ to
decay to top quarks nor for top quarks to decay into them. Furthermore, the
isosinglet component of neutral technipions may decay into a pair of gluons
if its constituent technifermions are colored. Thus, the decay
modes of interest to us are $\tpiz \ra \ol b b$ and, perhaps $\ol c c,
\tau^+\tau^-$, and $\tpipr \ra gg, \ts\ts \ol b b$. Branching
ratios are estimated from (for later use in the technihadron production
cross sections, we quote the energy-dependent widths~\cite{ellis,ehlq}):
\bea\label{eq:tpiwidths}
 \Gamma(\tpi \ra \ol f' f) &=& {1 \over {16\pi F^2_T}}
 \ts N_f \ts p_f \ts C^2_f (m_f + m_{f'})^2 \nn \\ \nn \\
 \Gamma(\tpipr \ra gg) &=& {1 \over {128 \pi^3 F^2_T}} 
 \ts \alpha^2_S \ts C_{\tpi} \ts \Ntc^2 \ts s^{{3\over{2}}} \ts .
\eea
Here, $C_f$ is an ETC-model dependent factor of order one {\it except} that
TC2 suggests $\vert C_t\vert \simle m_b/m_t$; $N_f$ is the number of colors
of fermion~$f$; $p_f$ is the fermion momentum; $\alpha_S$ is the QCD
coupling evaluated at $M_{\tpi}$; and $C_{\tpi}$ is a Clebsch of order one.
We take $M_{\tpi} = 110\,\gev$, $F_T \equiv \Few/3 = 82\,\gev$ for the
technipion decay constant (for nine isodoublets of technifermions),
$m_b = 4.2\,\gev$, $\alpha_S = 0.1$, $C_b = 1$ for $\tpiz$ and $\tpipr$,
and $C_{\tpi} = 4/3$. Then, the technipion partial widths are $\Gamma(\tpiz
\ra \ol b b) = \Gamma(\tpipr \ra \ol b b) = 35\,\mev$ and $\Gamma(\tpipr
\ra gg) = 10\,\mev$, quite narrow indeed.

As discussed in Refs.~\cite{multi,tpitev}, the standard two and
three technipion decay channels of the lightest $\troz$ and $\tom$ probably
are energetically forbidden. Then $\troz$ decays to $W^+_L W^-_L$ or
$W^\pm_L \tpimp$ and $\tom$ to $\gamma \tpiz$ or $Z^0 \tpiz$.
We parameterized this for $\tro$ decays with a simple model of two
isotriplets of technipions which are mixtures of $W_L^\pm$, $Z_L^0$ and
mass-eigenstate technipions $\tpipm$, $\tpiz$. The lighter isotriplet
$\tro$ is assumed to decay dominantly into pairs of the mixed state of
isotriplets $\vert\Pi_T\rangle = \sin\chi \ts \vert W_L\rangle + \cos\chi
\ts \vert\tpi\rangle$, where $\sin\chi = F_T/\Few$.
Then,
\be\label{eq:trhopipi}
\Gamma(\troz \ra \pi_A^+ \pi_B^-) = {2 \atro \CC^2_{AB}\over{3}} \ts
{\ts\ts p_{AB}^3\over {s}} \ts,
\ee
where $p_{AB}$ is the technipion momentum and $\atro$ is obtained by {\it
naive} scaling from the QCD coupling for $\rho \ra \pi\pi$, $\atro = 2.91
\ts (3/\Ntc)$. The parameter $\CC^2_{AB} = \sin^4\chi$ for $W_L^+ W_L^-$,
$\sin^2\chi \ts \cos^2\chi$ for $W_L^\pm \tpimp$, etc. The $\tro$ can be
very narrow: For $M_{\tro} = 210\,\gev$, $M_{\tpi} = 110\,\gev$,
and $\sin\chi = \third$, we have $\sum_{AB} \ts \Gamma(\troz \ra
\pi^+_A\pi^-_B) = 680\,\mev$, 80\% of which is $W_L^\pm \tpimp$.

We shall also need the decay rates of the $\tro$ to fermion-antifermion
states. These proceed through the $\troz$ coupling to $\gamma$ and $Z^0$:
\be\label{eq:trhoff}
\Gamma(\troz \ra \ol f_i f_i) = {N_f\ts \alpha^2 \over
{3\atro}} \ts {p_i \ts (s + 2m^2_i) \over {s}}\ts A^0_i(s) \ts.
\ee
Here, $\alpha$ is the fine-structure constant, $p_i$ is the momentum
and $m_i$ the mass of fermion $f_i$, and
\bea\label{eq:afactors}
A_i^0(s) &=& \vert \CA_{iL}(s) \vert^2
+ \vert \CA_{iR}(s) \vert^2 \ts, \nn \\
\CA_{i\lambda}(s) &=& Q_i + {2 \cos 2\thw \over {\sin^2 2\thw}} \ts
\zeta_{i \lambda} \left({s \over {s - M_Z^2 + i\ecm \ts
\Gamma_Z}}\right)\ts, \\
\zeta_{i L} &=& T_{3i} - Q_i \sin^2\thw, \qquad \zeta_{i R} = - Q_i
\sin^2\thw \ts. \nn
\eea
For parameters as above, the $\ol f f$
partial decay widths are $5.8\,\mev$ ($\ol u_i u_i$), $4.1\,\mev$ ($\ol 
d_i d_i$), $0.9\,\mev$ ($\ol \nu_i \nu_i$), and $2.6\,\mev$ ($\ell_i^+ 
\ell_i^-$).

For the $\tom$, phase space considerations suggest we consider only its
$\gamma \tpiz$ and fermionic decay modes. The energy-dependent widths are:
\bea\label{eq:tomegawidth}
\Gamma(\tom \ra \gamma \tpiz) &=& {\alpha p^3 \over {3 M_T^2}} \ts,
\nn \\ \nn \\
\Gamma(\tom \ra \ol f_i f_i) &=& {N_f \ts \alpha^2 \over
{3 \atro}} \ts {p_i\ts (s + 2m^2_i) \over {s}}\ts B^0_i(s) \ts.
\eea
The mass parameter $M_T$ in the $\tom \ra \gamma\tpiz$ rate is unknown
{\it a priori}; naive scaling from the QCD decay, $\omega \ra \gamma
\pi^0$, suggests it is several 100~GeV. The factor $B_i^0 =
\vert \CB_{iL} \vert^2 + \vert \CB_{iR} \vert^2$, where
\bea\label{eq:bfactors}
\CB_{i\lambda}(s) & = & 
\left[Q_i - {4 \sin^2\thw \over {\sin^2 2\thw}} \ts
\zeta_{i\lambda} \left({s \over {s - M_Z^2 + i\ecm\ts \Gamma_Z}}
\right)\right] \nn  \\ 
& & \times (Q_U + Q_D)\ts .
\eea
Here, $Q_U$ and $Q_D = Q_U - 1$ are the electric charges of the $\tom$'s
constituent technifermions. For $M_{\tom} = 210\,\gev$ and $M_{\tpi} =
110\,\gev$, and choosing $M_T = 100\,\gev$ and $Q_U = Q_D + 1 = {4\over
{3}}$, the $\tom$ partial widths are $115\,\mev$ ($\gamma \tpiz$),
$6.8\,\mev$ ($\ol u_i u_i$), $2.6\,\mev$ ($\ol d_i d_i$), $1.7\,\mev$ ($\ol
\nu_i \nu_i$), and $5.9\,\mev$ ($\ell_i^+ \ell_i^-$).

The beam momentum resolutions and corresponding annual integrated
luminosities of the First Muon Collider have been quoted to be $\sigma_p/p
= 3\times 10^{-5}$ ($\int \CL dt = 50\,\ipb$) for the narrow option at
$\ecm = 100\,\gev$ and $10^{-3}$ ($1\,\ifb$) at $\ecm =
200\,\gev$~\cite{deltap}. These correspond to beam energy spreads of
$\sigma_E \simeq 2\,\mev$ at 100~GeV and 150~MeV at 200~GeV. The
resolution at 100~GeV is less than the expected $\tpiz$, $\tpipr$ widths.
At 200~GeV it is sufficient to resolve the $\troz$, but not the $\tom$ for
the parameters we used. It is very desirable, therefore, that the 200~GeV
FMC's energy spread be 10~times smaller. Since each of these
technihadrons can be produced as an $s$-channel resonance, it would then be
possible to realize most of the theoretical peak cross section. These are
enormous, 2--3 orders of magnitude larger than the effective cross sections
that can be achieved at hadron and linear $e^+e^-$ colliders. To motivate
an improved resolution, we shall present results for $\sigma_p/p = 10^{-3}$
and $10^{-4}$ at $\ecm = 200\,\gev$, assuming in the latter case an annual
luminosity of only $0.1\,\ifb$.

Like the standard Higgs boson, neutral technipions are expected to couple
to $\mm$ with a strength proportional to $m_\mu$. Compared to $H^0$,
however, this coupling is enhanced by $\Few/F_T = 1/\sin\chi$.
This makes the FMC energy resolution well-matched to the $\tpiz$ width:
$\Gamma(\tpiz)/2\delta E \gg 1$ while $\Gamma(H^0)/2\delta E \simle 1$
Thus, the FMC is a technipion factory. Once a neutral technipion has been
found in $\tro$ or $\tom$ decays at a hadron collider, it should be
relatively easy to locate its precise position at the FMC. The cross
sections for $\ol f f$ and $gg$ production are isotropic; near the
resonance, they are given by
\bea\label{eq:tpirates}
& &{d\sigma(\mm \ra \tpiz \ts {\rm or} \ts \tpipr \ra \ol f f) \over{dz}} = 
\nn \\
& & \hspace{0.25in} {N_f \over {2\pi}} \ts \left({C_\mu C_f m_\mu m_f \over
 {F_T^2}}\right)^2 \ts
{s \over{(s - M_{\tpi}^2)^2 + s \ts \Gamma_{\tpi}^2}} \ts,
\nn \\ \\
& &{d\sigma(\mm \ra \tpipr \ra gg) \over{dz}} = \nn \\
& & \hspace{0.25in} {C_{\tpi} \over {32\pi^3}} \ts 
\left({C_\mu m_\mu\alpha_S\Ntc\over{F_T^2}}
\right)^2 \ts {s^2 \over{(s - M_{\tpi}^2)^2 + s \ts \Gamma_{\tpi}^2}}
\ts. \nn \\
\eea
Here, $z = \cos\theta$, where $\theta$ is the center-of-mass production
angle. For parameters as used below
Eq.~(\ref{eq:tpiwidths}), the theoretical peak cross sections are
$\sigma(\mm \ra \tpiz \ra \ol b b) = 1.4\,\nb$, $\sigma(\mm \ra \tpipr \ra
\ol b b) = 0.80\,\nb$, and $\sigma(\mm \ra \tpipr \ra gg) = 0.25\,\nb$. 
Angular cuts and $b$-detection efficiencies will decrease these rates.

\begin{figure}[tb]
\vbox to 8cm{
\vfill
\includegraphics{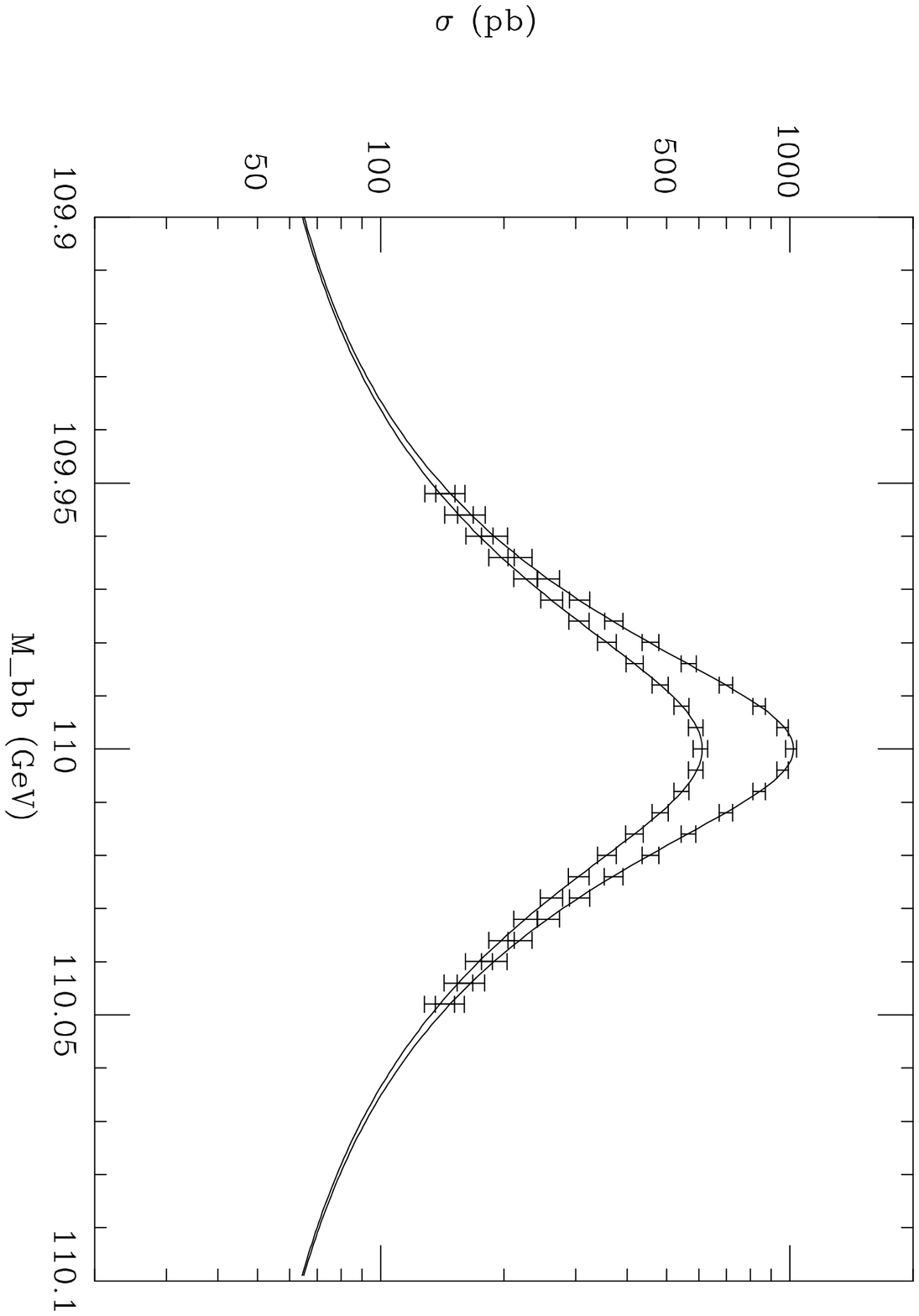}
\vfill
}
\caption{
Cross sections for $\mm \ra \tpiz \ra \ol b b$ (upper curve) and $\tpipr \ra
\ol b b$. Statistical errors only are shown for a luminosity of $1\,\ipb$
per point. Cuts and efficiencies are described in the text. The solid lines
are the theoretical cross sections  (perfect resolution).
\label{fig:one}}
\end{figure}

\begin{figure}[tb]
\vbox to 8cm{
\vfill
\includegraphics{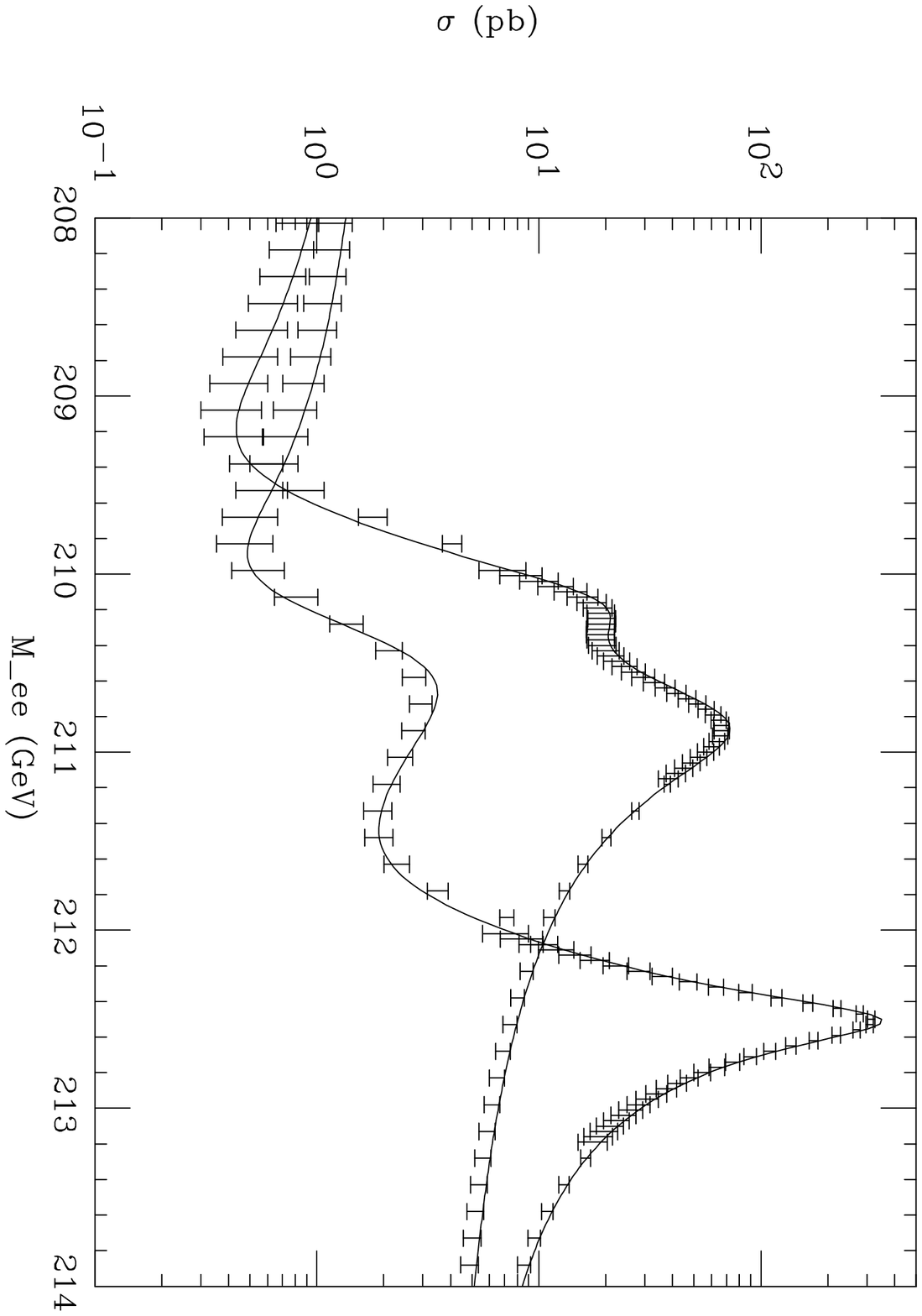}
\vfill
}
\caption{
Cross sections for $\mm \ra \tro, \ts \tom \ra e^+e^-$ for $M_{\rho_T} =
210\,\gev$ and $M_{\omega_T} = 211\,\gev$ (higher-peaked curve) and
$209\,\gev$. Statistical errors only are shown for resolutions and
luminosities described in the text. The solid lines are the theoretical
cross sections (perfect resolution).
\label{fig:two}}
\end{figure}

In Fig.~1 we show the $\tpiz$ and $\tpipr \ra \ol b b$ signals and
$\gamma$, $Z^0$ background for $\delta E = 2\,\mev$ and an integrated
luminosity of only $25\,\ipb$. We have assumed $\vert \cos\theta \vert <
0.95$ and a single $b$-tag efficiency of 50\%. The peak cross sections are
$1.0\,\nb$ and $0.6\,\nb$, respectively, over a background of $65\,\pb$.
Statistical errors only are shown. It is obvious that the widths of these
resonances can be distinguished from one another. We have not considered
the interesting and likely possibility of $\tpiz$--$\tpipr$ interference.
Such interferences are examined below for $\tro$ and $\tom$. The process
$\tpipr \ra gg$, not shown here, has a signal to ($\ol q q$) background of
$250/250\,\pb$ and can be used to determine which resonance (or mixture) is
being observed. Note that this channel will not show up in a heavy-flavor
tag. Furthermore, we do not expect a $\ol UU$ technipion to decay to $\ol b
b$. We conclude that the FMC can carry out very precise studies of the
neutral $\tpi$ unless they are nearly degenerate with the $Z^0$.

A small nonzero isospin splitting between $\troz$ and $\tom$ would appear
as a dramatic interference in the $\mm \ra \ol f f$ cross section {\it
provided} the FMC energy resolution is good enough. The cross section is
calculated by using the full $\gamma$--$Z^0$--$\tro$--$\tom$ propagator
matrix, $\Delta(s)$. With $\CM^2_V = M^2_V - i \ecm \ts \Gamma_V(s)$ for $V
= Z^0,\tro,\tom$, this matrix is the inverse of
\be\label{eq:vprop}
\Delta^{-1}(s) =\left(\ba{cccc}
s & 0 & -s f_{\gamma\tro} & -s f_{\gamma\tom} \\
0 & s - \CM^2_Z  & -s f_{Z\tro} & -s f_{Z\tom} \\
-s f_{\gamma\tro}  & -s f_{Z\tro}  & s - \CM^2_{\tro} & 0 \\
-s f_{\gamma\tom}  & -s f_{Z\tom}  & 0 & s - \CM^2_{\tom} 
\ea\right) \ts.
\ee
Here, $f_{\gamma\tro} = \xi$, $f_{\gamma\tom} = \xi \ts (Q_U + Q_D)$,
$f_{Z\tro} = \xi \ts \tan 2\thw$, and $f_{Z\tom} = - \xi \ts
\sin^2\thw/\sin 2\thw \ts (Q_U + Q_D)$, where $\xi = \sqrt{\alpha/\atro}$.
The cross section is given in terms of matrix elements of $\Delta$ by
\bea\label{eq:mmffrate}
& &{d\sigma(\mm \ra \troz,\ts \tom \ra \ol f_i f_i) \over{dz}} = \nn \\
& &\hspace{0.25in}{N_f \pi \alpha^2\over{8s}} \biggl\{
\left(\vert\CD_{iLL}\vert^2 + \vert\CD_{iRR}\vert^2\right)(1+z)^2 \nn \\
& &\hspace{0.25in} +\left(\vert\CD_{iLR}\vert^2 + \vert\CD_{iRL}\vert^2\right)(
1-z)^2
\biggr\} \ts;
\eea
where
\bea\label{eq:dfactors}
\CD_{i\lambda\lambda'}(s) &=& s\biggl[Q_i Q_\mu \ts \Delta_{\gamma\gamma}(s)
 + {4\over{\sin^2 2\thw}} \ts \zeta_{i \lambda}
\ts \zeta_{\mu \lambda'} \ts \Delta_{ZZ}(s) \nn \\
&& + {2\over{\sin 2\thw}} \ts \biggl(\zeta_{i \lambda} Q_\mu
\Delta_{Z\gamma}(s) + Q_i \zeta_{\mu \lambda'} \Delta_{\gamma Z}(s)\biggr)
\biggr]
\ts. \nn \\
\eea

Figure~2 shows the interference effects in $\mm \ra e^+e^-$ for input
masses $M_{\tro} = 210\,\gev$ and $M_{\tom} = 209$ and $211\,\gev$. It is
assumed that the resonance region (first isolated in a hadron collider) is
scanned in 40~steps with a $1\,\ifb$ run at coarse resolution, $\delta E =
150\,\mev$. The resonances are then studied with $\delta E = 15\,\mev$ in
a $100\,\ipb$ run with forty $30\,\mev$ wide steps. As before, $\vert \cos
\theta \vert < 0.95$. Because of the precise FMC beam energies, this is
just a counting experiment and does not require excellent $e^\pm$ energy
measurement. The same applies to $\ol q q$ final states. The effect of
changing the $\tro$--$\tom$ mass difference by 2~GeV is striking. In both
cases shown, the $\tro$ is the broader structure peaking near
$210.8\,\gev$. For input $M_{\omega_T} = 209\,\gev$, the narrow resolution
picks $\tom$ out as the flat shoulder at $210.2\,\gev$. The dip is a
somewhat more pronounced in $\ol q q$ final states. For input $M_{\omega_T}
= 211\,\gev$, narrow resolution reveals a majestic peak at $212.5\,\gev$
with $\sigma(\mm \ra e^+e^-) = 325\,\pb$. This demonstrates the importance
of precise resolution in the $200\,\gev$ muon collider.

Large cross sections such as these, plus the ability to measure $e^\pm$ 
charges, make possible detailed angular distribution measurements. These 
will be even more incisive if the muon beams can be polarized without 
great loss in luminosity. These features of the FMC will be essential for 
studying the charges and isospins that appear in Eqs.~(\ref{eq:afactors})
and (\ref{eq:bfactors}).

Before closing, we mention that associated production of technipions with
weak bosons also occurs at very large rates (see Ref.~\cite{tpitev} for the
cross section formulae). For the parameters used above, $\sigma(\mm \ra
\troz \ra W_L^\pm \tpimp) = 0.9\,\nb$ and $\sigma(\mm \ra \tom \ra \gamma
\tpiz) = 8.9\,\nb$. This offers an unparalleled opportunity to study 
charged technipion decay processes in a relatively clean setting.

To sum up: modern technicolor models predict narrow neutral technihadrons,
$\tpi$, $\tro$ and $\tom$. These states would appear as spectacular,
high-rate resonances in a $\mm$ collider with $\ecm = 100$--$200\,\gev$ and
energy resolution $\sigma_E/E \simle 10^{-4}$. This is a very strong
physics motivation for building the First Muon Collider.

We thank members of the FMC Workshop Strong Dynamics
Subgroup for valuable interactions, especially P.~Mackenzie and C.~Hill for
stressing the importance of a narrow $\tpiz$ at a muon collider, and
P.~Bhat for assistance and encouragement. We also thank S.~Geer and
R.~Palmer for discussions on the FMC parameters. We are indebted to
T.~Sjostrand for first pointing out to us the likely importance of the
$\troz$ and $\tom$ decays to fermions. The research of EE and JW is
supported by the Fermi National Accelerator Laboratory, which is operated
by Universities Research Association, Inc., under
Contract~No.~DE--AC02--76CHO3000. KL's research is supported in part by the
Department of Energy under Grant~No.~DE--FG02--91ER40676. KL thanks
Fermilab for its hospitality during various stages of this work.


\end{document}